\documentclass[letterpaper, 10 pt, conference]{ieeeconf}

\IEEEoverridecommandlockouts                              

\usepackage{nomencl}

\usepackage{amsthm}

\makeglossary
\usepackage{graphicx}
\usepackage{caption}
\usepackage{subcaption}
\usepackage{algorithm}
\usepackage{algorithmic}
\usepackage{amsmath}
\usepackage{graphicx}
\usepackage{enumerate}%
\usepackage{type1cm}
\usepackage{lettrine}
\usepackage{breqn}
\usepackage{graphicx}
\usepackage{amssymb}
\usepackage{framed}
\usepackage{tikz}
\usepackage{amsmath}
\usepackage{setspace}
\usepackage{pgfplots}
\usepackage{caption}
\usepackage{subcaption}

\usepackage{upgreek}

\usepackage{epstopdf}

\usepackage{bm}

\usepackage{bbm}

\newcommand{\diag}{\mathop{\rm diag}}

\newtheorem{problem}{\textnormal{\textbf{Problem}}}

\newtheorem{assumption}{Assumption}

\newtheorem{lemma}{\textbf{Lemma}}
\newtheorem{corollary}{Corollary}

\usepackage{calc}

\newtheorem{theorem}{\textnormal{\textbf{Theorem}}}

\newcommand{\nunder}[2][5]{\mathrlap{\mkern\the\numexpr#1/2mu\relax\underlipe{\phantom{\mathrm{#2}\mkern-#1mu}}}#2}
\usepackage[font={footnotesize}]{caption}

\title{\LARGE \bf  Stability Analysis of Droop-Controlled Inverter-Based Power Grids via Timescale Separation
}

\begin{document}
	\author{Stefanos Baros, Christoforos N. Hadjicostis, and  Francis O’Sullivan
\thanks{S. Baros is with the National Renewable Energy Laboratory, Golden, CO 80401, USA,
stefanos.baros@nrel.gov. This work was conducted while he was a postdoctoral researcher at MIT Energy Initiative, Cambridge, MA.}
\thanks{Christoforos Hadjicostis is with the ECE Department of University of Cyprus, Nicosia, Cyprus, chadjic@ucy.ac.cy.}
\thanks{Francis O' Sullivan was with the MIT Energy Initiative and is now with Orsted Onshore North America, frankie@mit.edu.}}

	\maketitle
	\thispagestyle{empty}
	\pagestyle{empty}

	\begin{abstract}
		We consider the problem of stability analysis for distribution grids with droop-controlled inverters and dynamic distribution power lines. The inverters are modeled as voltage sources with controllable frequency and amplitude. This problem is very challenging for large networks as numerical simulations and detailed eigenvalue analysis are impactical. Motivated by the above limitations, we present in this paper a systematic and computationally efficient framework for stability analysis of inverter-based distribution grids. To design our framework, we use tools from singular perturbation and Lyapunov theories.  Interestingly, we show that stability of the fast dynamics of the power grid depends only on the voltage droop gains of the inverters while,  stability of the slow dynamics, depends on both voltage and frequency droop gains. Finally, by leveraging these timescale separation properties, we derive sufficient conditions on the frequency and voltage droop gains of the inverters that warrant stability of the full system. We illustrate our theoretical results through a numerical example on the IEEE 13-bus distribution grid.

	\end{abstract}

	\IEEEpeerreviewmaketitle

	\section{Introduction}
	\par Enviromental and economic reasons together with recent technological advances are the primary drivers for high penetration of renewable energy resources (RERs) in  power systems \cite{Baros_NG_reliability}, \cite{Baros_NG_conf}, \cite{Baros_DRMM}, \cite{Baros_emerging_control}. Today, renewable energy is mainly generated in distribution grids closer to the end-users by small distributed energy resources (DERs) that are interfaced with the main grid through alternating current (AC) inverters \cite{pogaku}.  When the penetration levels are relatively small, the power generation of RERs can be often treated as negative demand and the control objective for RERs is maximum power output generation \cite{Baros_dist_torque_control}.  In such scenarios, the employable control techniques for AC inverters are limited to maximum power point tracking strategies \cite{Baros_dist_torque_control}. However, as the penetration of RERs around the world increases, these control techniques are not offered anymore as viable options \cite{Baros_dist_torque_control}, \cite{Baros_dist_stor_control}.  It is imperative that RERs use control methods for their inverters that enable them to attain better regulation of their power outputs \cite{Baros_dist_torque_control}, \cite{Baros_dist_consensus_optimal}, \cite{Baros_dist_stor_control}. One such control method that emerged in the early `90s is droop control \cite{droopcontrolfirstpaper}. Droop control can allow inverter-interfaced RERs to satisfy a given load demand while achieving a certain power sharing distribution in steady-state.
	\par The stability analysis problem for droop-controlled inverter-based grids was first investigated in \cite{stabilityinverters}, \cite{guerrerostabilityfirst} and \cite{pogaku} using detailed small-signal analysis. This kind of analysis usually involves the repeated computation of the system's eigenvalues and extensive numerical simulations. In the same spirit,  stability of droop-controlled inverter-based power grids  was studied more recently in \cite{hatzistab}, \cite{guerrerostab2} and \cite{guerrerostability} via sensitivity and eigenvalue analysis. These approaches are easily implementable but they can be computationally very costly  as the network size increases. This is widely recognized by the scientific community and recently various researchers focused on examining stability of droop-controlled inverter-based grids via Lyapunov-like approaches. Representative publications along this line of work are \cite{turitsyn2}, \cite{Schiffer} and \cite{dorfler}. In \cite{turitsyn2}, an approach for obtaining a reduced-order model for a droop-controlled inverter-based microgrid with electromagnetic network dynamics was first presented. A Lyapunov function for the reduced-order system was then constructed that led to decentralized sufficient stability conditions.   In \cite{Schiffer}, sufficient conditions for stability of meshed microgrids with droop-controlled inverters were derived. As shown and explained in \cite{turitsyn2} however, it is absolutely necessary to consider the network dynamics in the stability analysis of droop-controlled inverter-based power grids as these may greatly impact the obtained stability regions. Finally, in \cite{dorfler}, stability of inverter-based power grids with grid-forming virtual oscillator control and line dynamics was studied, using a Lyapunov-like approach and singular perturbation theory. 
	\par \textbf{\textit{Contributions}}. In this paper, we study the stability problem for distribution grids composed of constant impedance loads, droop-controlled inverters and dynamic distribution lines. We model inverters as voltage sources with controllable amplitude and frequency. Contrary to \cite{Schiffer}, we do not require constant voltage amplitudes and incorporate line dynamics in our analysis.  Our main contribution is two-fold. We first present a systematic framework for conducting stability analysis of inverter-based power grids in a compositional and computationally efficient fashion. Subsequently, we apply our proposed framework and derive simple sufficient stability conditions.  Contrary to \cite{turitsyn2}, when our derived conditions hold they result in guaranteed stability of the original full system and not only of the low-order system approximation. 
	\par The remainder of the paper is structured as follows. In Section II, we review the dynamical model of a droop-controlled inverter-based power grid with line dynamics. In Section III, we formulate the stability problem. In Section IV we present our main results. In Section V, we verify our results through a numerical example. Finally in Section VI, we conclude this paper.
	\section{Review of Inverter-based Power Grid Model}
	We consider a distribution grid comprised of $N$ droop-controlled inverters, $M$ transmission lines and $L$ loads, described by the sets $\mathcal{N}, \mathcal{E}$ and $\mathcal{L}$, respectively.
	\subsection{Droop-controlled Inverter-based Power Grid}
	We model inverters as AC voltage sources with controllable amplitude and frequency, lines as $``R-L"$  circuits, and loads as constant impedances. Without loss of generality, we assume that loads exist only on inverter buses \cite{turitsyn2}, \cite{Schiffer}.
	\par The state-variables associated with the inverters are the angles, frequencies and amplitudes of the output voltages denoted by, $\delta:=(\delta_1,...,\delta_N)^T\in\mathbb{R}^N,\;\omega:=(\omega_1,...,\omega_N)^T\in\mathbb{R}^N$ and $V:=(V_1,...,V_N)^T\in\mathbb{R}^N$, respectively. The state-variables associated with the transmission lines are the currents $I_D:=(I_{D,1},...,I_{D,M})^T\in\mathbb{R}^M$ and $I_Q:=(I_{Q,1},...,I_{Q,M})^T\in\mathbb{R}^M$, expressed in a $dq$ reference frame. We note that, $\delta\in\mathbb{R}^N$ are measured in radians, $\omega\in\mathbb{R}^N$ in $(r/s)$ while $V\in\mathbb{R}^N$, $I_D\in\mathbb{R}^M, I_Q\in\mathbb{R}^M$ and $P, Q\in\mathbb{R}^N$ are measured in per unit values. The matrices $R=\diag(\{R_{i}\}_{i=1}^M)$, $X=\diag(\{X_{i}\}_{i=1}^M)$ and $L=\diag(\{L_{i}\}_{i=1}^M)$ are all diagonal matrices with the per unit resistances, reactances, and inductances on their diagonals, respectively.  The constant vectors $\omega_d:=(\omega_{d,1},...,\omega_{d,N})^T\in\mathbb{R}^N$, $V_d:=(V_{d,1},...,V_{d,N})^T\in\mathbb{R}^N$, $P_d:=(P_{d,1},...,P_{d,N})^T\in\mathbb{R}^N$ and $Q_d:=(Q_{d,1},...,Q_{d,N})^T\in\mathbb{R}^N$ collect respectively, the desired frequencies, voltage amplitudes and the real and reactive power outputs of the inverters.  $V_b$ and $S_b$  are the base voltage and base power, respectively. Finally, $K_P:=K_P^{pu}\omega_b=(N_PS_b),\;K_P:=(\diag\{k_{P,i}\}_{i=1}^N)$ are the scaled frequency droop gains that map per unit power output changes to frequency changes in $(r/s)$ and $K_Q:=K_Q^{pu}=K_Q(S_b/V_b),\;\;K_Q:=(\diag\{k_{Q,i}\}_{i=1}^N)$ are the per unit voltage droop gains.   
	Given the above notation, the model of a distribution grid can be expressed compactly in state-space form \cite{turitsyn2}, \cite{Schiffer}
\begin{align}
\frac{d\delta}{dt}&=(\omega-\omega_b\mathbf{1}_N)\label{pumodel1}\\
\frac{d\omega}{dt}&=T_P^{-1}[-\omega+\omega_d-K_P(P-P_d)]\label{pumodel2}\\
\frac{dV}{dt}&=T_Q^{-1}[-V+V_d-K_Q(Q-Q_d)]\label{pumodel3}\\
\frac{dI_D}{dt}&=\omega_bL^{-1}[-RI_D+XI_Q+C^{\mathcal{T}} \overline{V}_D] \label{pumodel4}\\
\frac{dI_Q}{dt}&=\omega_bL^{-1}[-RI_Q-XI_D+C^{\mathcal{T}} \overline{V}_Q] \label{pumodel5}
\end{align}
 The $d$-$q$ components of the extended voltage vector $\overline{V}_D,\; \overline{V}_Q$ are given by
	\begin{align}
	\overline{V}_D=\Big((\diag(\{\cos(\delta_i)\}_{i=1}^N)V)^T, \;\;\; V_{g,D} \Big)^T\in\mathbb{R}^{N+1}\label{VDbar}\\
	\overline{V}_Q=\Big((\diag(\{\sin(\delta_i)\}_{i=1}^N)V)^T, \;\;\; V_{g,Q} \Big)^T\in\mathbb{R}^{N+1}\label{VQbar}
	\end{align}
	where $V_{g,D},\;V_{g,Q}\in\mathbb{R}$ are the $d$-$q$ components of the feeder's voltage and $\omega_b$ is the grid's nominal frequency $2\pi60 \;(r/s)$. $T_P=\diag(\{T_{P,i}\}_{i=1}^N)$ and $T_Q=\diag(\{T_{Q,i}\}_{i=1}^N)$ are the time constants of the low-pass filters.   The real power of the inverters $P\in\mathbb{R}^N$ can be expressed as
	\begin{align}
	P&=\diag(\{\cos(\delta_i)\}_{i=1}^N)\diag(\{V_i\}_{i=1}^N)[C^{\mathcal{I}}I_D+C^{\mathcal{LD}}V_{DQ}]\nonumber\\
	&+\diag(\{\sin(\delta_i)\}_{i=1}^N)\diag(\{V_i\}_{i=1}^N)[C^{\mathcal{I}}I_Q+C^{\mathcal{LQ}}V_{DQ}]\nonumber
	\end{align}
	while the reactive power output $Q\in\mathbb{R}^N$ as
	\begin{align}
	Q&=\diag(\{\sin(\delta_i)\}_{i=1}^N)\diag(\{V_i\}_{i=1}^N)[C^{\mathcal{I}}I_D+C^{\mathcal{LD}}V_{DQ}]\nonumber\\
	&-\diag(\{\cos(\delta_i)\}_{i=1}^N)\diag(\{V_i\}_{i=1}^N)[C^{\mathcal{I}}I_Q+C^{\mathcal{LQ}}V_{DQ}]\nonumber
	\end{align}
 The vector $V_{DQ}$ is defined as
	\begin{align}
	V_{DQ}=\Big((\diag(\{\cos(\delta_i)\}_{i=1}^N)V)^T, (\diag(\{\sin(\delta_i)\}_{i=1}^N)V)^T \Big)^T \nonumber
	\end{align}
	The load impedance matrices $C^{\mathcal{LD}}:=\{c^{\mathcal{LD}}_{ij}\}\in\mathbb{R}^{N\times 2N}$ and $C^{\mathcal{LQ}}:=\{c^{\mathcal{LQ}}_{ij}\}\in\mathbb{R}^{N\times 2N}$ are defined as
	\begin{align}
	c^{\mathcal{LD}}_{ij}=\begin{cases}\frac{R_{L,k}}{R_{L,k}^2+X_{L,k}^2},   \text{ if load } k \text{ lies at inv. bus } i \text{ and } j=i \\
	\frac{X_{L,k}}{R_{L,k}^2+X_{L,k}^2}, \text{ if load } k \text{ lies at inv. bus } i \text{ and } j=i+N\\
	0, \text{ \hspace{13mm} otherwise }\end{cases}\nonumber
	\end{align}
	\begin{align}
	c^{\mathcal{LQ}}_{ij}=\begin{cases}\frac{-X_{L,k}}{R_{L,k}^2+X_{L,k}^2}, \text{ if load } k \text{ lies at inv. bus } i \text{ and } j=i  \\
	\frac{R_{L,k}}{R_{L,k}^2+X_{L,k}^2},    \text{ if load } k \text{ lies at inv. bus } i \text{ and } j=i+N \\
	0, \text{ \hspace{13mm} otherwise }\end{cases}\nonumber
	\end{align}
	Above, $R_{L,k}$ and $X_{L,k}$ represent the resistance and reactance respectively of load $k$, where $k\in\mathcal{L}$. Further,  $C^{\mathcal{I}}:=\{c^{\mathcal{I}}_{ij}\}\in\mathbb{R}^{N\times M}$ is the incidence matrix, the entries of which are defined as
	\begin{align}
	c^{\mathcal{I}}_{ij}=\begin{cases}-1, \hspace{4mm} \text{ if inverter } i \text{ lies at the end of line } j\\
	1, \hspace{5mm} \text{  \hspace{1mm} if inverter } i \text{ lies at the beginning of line } j\\
	0, \text{ \hspace{6mm} otherwise }\end{cases}\nonumber
	\end{align}
	In addition, $C^{\mathcal{T}}:=\{c^{\mathcal{T}}_{ij}\}\in\mathbb{R}^{M\times (N+1)}$ is another incidence matrix, the entries of which can be defined as
	\begin{align}
	c^{\mathcal{T}}_{ij}=\begin{cases}-1, \hspace{5mm} \text{ if inverter } j \text{ lies at the end of line } i\\
	1, \hspace{8mm} \text{   if inverter } j \text{ lies at the beginning of line } i\\
	0, \text{  \hspace{7mm}            otherwise }\end{cases} \nonumber
	\end{align}
	In defining $C^{\mathcal{I}}$, $C^{\mathcal{T}}$ we used the following convention. For a transmission line $j:=(a,b)\in\mathcal{E}$, we consider $a$ to be its beginning and $b$ its end when the current is flowing from node $a$ to node $b$. 

	\section{Problem Formulation}
 Despite the wide use of droop-based control methods, today there is still a lack of systematic methodologies for stability analysis of power grids that accommodate numerous droop-controlled inverters. It is also not clear how we should best approach the stability analysis problem for such systems, in order to characterize the ranges for the droop control gains of the inverters that lead to guaranteed stability \cite{turitsyn2}, \cite{Schiffer}. In light of this, the main research problem still remains open and can be formulated as follows.
	\begin{problem}
		\par   Consider the system $\eqref{pumodel1}$--$\eqref{pumodel5}$ linearized around an equilibrium point $w_0$
		\begin{align}
		\frac{dw(t)}{dt}=Aw(t),\;\;w(t)\in\mathbb{R}^{3N+2M} \label{linearsystem}
		\end{align}
		where $w=[\delta^\top,\;\omega^\top,\;V^\top,\; I_D^\top,\; I_Q^\top]^\top$ and $f$ is the vector field of  $\eqref{pumodel1}$--$\eqref{pumodel5}$ and $A:=A(K_P,K_Q)$ is the Jacobian matrix of $f$. Develop a systematic and computationally efficient framework for tuning the frequency droop gains  $K_P$ and voltage droop gains $K_Q$ so that asymptotic stability of the equilibrium of \eqref{linearsystem} is guaranteed.\label{problem1}
	\end{problem}	
	The rest of the paper is devoted to addressing Problem~\ref{problem1}.
	\section{Main Results}
	In this section, we present our results on the stability problem for droop-controlled inverter-based power grids. Due to space limitation, we only provide (in the Appendix) the proofs of the main results in this section.
	\subsection{Inverter-based Power Grid in Multi-parameter Singularly Perturbed Form}
	\par We begin by making the reasonable assumption that the time-constants of the low-pass filters are the same.
	\begin{assumption}[Homogeneous filter time constants]
		\begin{align}
		T_{P,1}=...=T_{P,N}=T_{Q,1}=...=T_{Q,N}=\varepsilon_1\in\mathbb{R}_+
		\end{align}\label{assum1}
	\end{assumption}
\vspace{-6mm}
	Under Assumption~\ref{assum1}, the linearized dynamics  of an inverter-based power grid described by \eqref{linearsystem} can be written in the  \textit{multi-parameter singularly perturbed} form \cite{siljakmultiparameter}, \cite{Kokotovic}
	\begin{align}
	\frac{d x}{dt}&=A_{xz}z\label{inv1}\\
	\varepsilon_1 \frac{dz}{dt}&= A_{zz}z+A_{zx}x+A_{zy}y\label{inv2}\\
	E_2 \frac{dy}{dt}&= A_{yz}z+A_{yx}x+A_{yy}y\label{inv3}
	\end{align}
	where the state-vector $w$ can be broken down to:
	\begin{align}
	x&=[\Delta\delta_1,...,\Delta\delta_{N}]^T\in\mathbb{R}^{N}\\
	z&=[\Delta\omega_1,...,\Delta\omega_{N}, \Delta V_1,...,\Delta V_N]^T\in\mathbb{R}^{2N}\\
	y&=[\Delta I_{D,1},...,\Delta I_{D,M} , \Delta I_{Q,1},...,\Delta I_{Q,M}]^T\in\mathbb{R}^{2M}
	\end{align}
We note that, all the matrices that appear in \eqref{inv1}--\eqref{inv3} can be obtained appropriately from the Jacobian matrix $A$.	The matrix $E_2\in\mathbb{R}^{2M\times 2M}$ can be compactly expressed as
	\begin{align}
	E_2:=\begin{bmatrix}\diag(\{\varepsilon_{2,i}\}_{i=1}^{M}) & 0_{M\times M}\\ 0_{M\times M} &\diag(\{\varepsilon_{2,i}\}_{i=1}^{M})\end{bmatrix}
	\end{align}
	where $\varepsilon_{2,i}:=(L_i/\omega_b)$. 	It is evident that the dynamic behavior of \eqref{inv1}-\eqref{inv3} is dictated by $(M+2)$ distinct time scales $t, t_z:=(t/\varepsilon_1)$ and $t_{y,i}:=(t/\varepsilon_{2,i})$ where $i\in\mathcal{E}$.
	\subsection{From Multi-parameter to Standard Singularly Perturbed Form}
We assume that the values of the line inductances $\varepsilon_{2,i}$ are all of the same order and this order is significantly different from the order of the time constant of the filters $\varepsilon_1$. This assumption allows us to specify $\varepsilon_2:=(\varepsilon_{2,1}\cdot\cdot\cdot \varepsilon_{2,M})^{1/M}$ and bring the inverter-based power grid \eqref{inv1}-\eqref{inv3} to the more convenient \textit{standard singularly perturbed form} \cite{siljakmultiparameter}, \cite{Kokotovic}
	\begin{align}
	\frac{d x}{dt}&=A_{xz}z\label{inv1_3}\\
	\varepsilon_1 \frac{dz}{dt}&= A_{zz}z+A_{zx}x+A_{zy}y\label{inv2_3}\\
	\varepsilon_2 \frac{dy}{dt}&= D_2(A_{yz}z+A_{yx}x+A_{yy}y)\label{inv3_3}
	\end{align}
	where the new matrix $D_2$ is given by
	\begin{align}
	D_2:=\begin{bmatrix}\diag(\{\varepsilon_2/\varepsilon_{2,i}\}_{i=1}^{M}) & 0_{M\times M}\\ 0_{M\times M} &\diag(\{\varepsilon_2/\varepsilon_{2,i}\}_{i=1}^{M})\end{bmatrix}
	\end{align}
	It should be now obvious that the dynamic behavior of the new system \eqref{inv1_3}--\eqref{inv3_3} is dictated by three timescales  $t, t_z:=(t/\varepsilon_1)$ and $t_{y}:=(t/\varepsilon_{2})$ with the small parameters  $\varepsilon_1$ and $\varepsilon_2$ giving rise to these timescales.  We designate $x,z$ and $y$ to be the \textit{slow}, \textit{fast} and \textit{very-fast} state-variables, respectively.
	\subsection{Suppressing the Electromagnetic Network Dynamics}
	\par Our goal here is to derive a reduced-order model for the inverter-based power grid \eqref{inv1_3}--\eqref{inv3_3} by suppressing the network dynamics. We start by expressing \eqref{inv1_3}--\eqref{inv3_3} with respect to $t_z=(t/\varepsilon_1)$ in the form \cite{Kokotovic}
	\begin{align}
	\frac{d x}{dt_z}&=A_{xz}\varepsilon_1 z\label{inv1_z}\\
	\frac{dz}{dt_z}&= A_{zz}z+A_{zx}x+A_{zy}y\label{inv2_z}\\
	\varepsilon_3 \frac{dy}{dt_z}&= D_2(A_{yz}z+A_{yx}x+A_{yy}y)\label{inv3_z}
	\end{align}
	where $\varepsilon_{3}:=(\varepsilon_{2}/\varepsilon_1)$. By writing the system in this form, we uncover the timescale separation between the dynamics of the inverters' states $(x,z)$ and the network's states $y$. 
	\subsection{Very-fast Boundary-layer Subsystem}
	Our next task is to derive the very-fast boundary-layer electromagnetic dynamics of the network and investigate their stability properties. To do that, we first compute the zeroth-order manifold of $y$,  $y_0(x,z)=A_{0z}z+A_{0x}x \label{y0}$	where $A_{0z}:=(-A_{yy}^{-1}A_{yz})$, $A_{0x}:=(-A_{yy}^{-1}A_{yx})$ and $A_{yy}=\begin{pmatrix}-R & X\\-X & -R\end{pmatrix}$. It is easy to notice that $A_{yy}$ is always invertible.  We perform change of variables $\xi:=(y-y_0)$ to obtain from \eqref{inv3_z}
	\begin{align}
	\frac{d\xi}{d\tau}=\tilde{A}_{yy}\xi,\hspace{7mm}\xi\in\mathbb{R}^{2M} \label{veryfastsystem}
	\end{align}
	where $\tau=(t_z/\varepsilon_3)$, $\tilde{A}_{yy}:=(D_2A_{yy})$. 	Practically, \eqref{veryfastsystem} describes the trajectories of the network's currents when seen decoupled from the much slower dynamics of the output voltages of the inverters. 
	\subsection{Stability of Very-fast Boundary-layer Subsystem}
	At this point, we establish stability of the equilibrium of \eqref{veryfastsystem} through the following lemma.
	\begin{lemma}
		The equilibrium $\xi^*=0_{2M}$ of the very-fast boundary layer subsystem \eqref{veryfastsystem} is asymptotically stable.
		\label{lemma1}
	\end{lemma}	
 Intuitively, Lemma \ref{lemma1} says that the deviations of the network's currents $\xi$ converge to  $\xi^*=0_{2M}$, as $\tau\to \infty$, when the inverter's states  $x,z$ are ``frozen''.  The following corollary ensues from Lemma \ref{lemma1}.
 \begin{corollary}
 	Let the matrix $\tilde{A}_{yy}$ describing the decoupled transmission line dynamics be Hurwitz. Then, $W:=\xi^TP_{\xi} \xi$ with $P_{\xi}\succ 0$  is a Lyapunov function for the very fast boundary-layer system \eqref{veryfastsystem} satisfying
 	\begin{align}
 	\frac{\partial W}{\partial \xi}\mu(x,\eta,\xi+y_0)&\leq -\alpha_4 \psi_4^2(\xi)\label{thm2bound2}\\
 	-\frac{\partial W}{\partial \xi}\begin{bmatrix}\frac{\partial y_0}{\partial x} & \frac{\partial y_0}{\partial \eta} \end{bmatrix}h(x,\eta,\xi+y_0)&\leq\nonumber\\ &\beta_4\psi_3(x,\eta)\psi_4(\xi)+\gamma_2\psi_4^2(\xi)\label{thm2bound4}
 	\end{align}
 	where  $\alpha_4:=\lambda_{min}(Q_{\xi})$ with $Q_{\xi}:=(\tilde{A}_{yy}^TP_{\xi}+P_{\xi}\tilde{A}_{yy})$. Further, $\psi_4(\xi):=\|\xi\|_2$ and $\psi_3(x,\eta):=\|\psi_1(x), \psi_2(\eta)\|_2$  with $\beta_4:=\max(\sigma_{max}(\Theta_{\xi x}),\sigma_{max}(\Theta_{\xi \eta}))\sqrt{2}$ and $\gamma_2:=-\lambda_{min}(\Theta_{\xi\xi})$, where
 	\begin{align}
 	\Theta_{\xi x}&:=	(P_\xi+P_{\xi}^T)\Big[(A_{0x}+A_{0z}\Gamma_0)A_{xz}\varepsilon_1\Gamma_0\nonumber\\
 	&+A_{0z}(A_{zz}\Gamma_0+A_{zx}-\Gamma_0A_{xz}\varepsilon_1\Gamma_0)\nonumber\\
 	&+A_{0z}A_{zy}(A_{0x}+A_{0z}\Gamma_0))  \Big]\label{thetaxichi}\\
 	\Theta_{\xi\eta}&:=(P_\xi+P_{\xi}^T)\Big[(A_{0x}+A_{0z}\Gamma_0)A_{xz}\varepsilon_1\nonumber\\
 	&+A_{0z}(A_{zz}-\Gamma_0A_{xz}\varepsilon_1+A_{zy}A_{0z})\Big]\label{thetaxieta}\\
 	\Theta_{\xi \xi}&:=(P_{\xi}+P_{\xi}^T)A_{0z}A_{zy} \label{thetaxixi}
 	\end{align}
 	\label{corollary3}
 \end{corollary}	
In the sequel, we examine the stability properties of the reduced-order power grid, obtained by suppressing the electromagnetic network dynamics.
		\subsection{Reduced-order Model of Inverter-based Power Grid}
The reduced-order power grid model can be obtained from \eqref{inv1}--\eqref{inv2} upon substitution of the very-fast manifold $y_0$ \cite{Kokotovic}
	\begin{align}
	\frac{dx}{dt}&=A_{xz}z \label{inv1_red}\\
	\varepsilon_1\frac{dz}{dt}&=\tilde{A}_{zz}z+\tilde{A}_{zx}x \label{inv2_red}
	\end{align} 
	where $\tilde{A}_{zz}:=(A_{zz}+A_{zy}A_{0z})\in\mathbb{R}^{2N\times 2N}$ and $\tilde{A}_{zx}:=(A_{zx}+A_{zy}A_{0x})\in\mathbb{R}^{2N\times N}$. As the matrices $\tilde{A}_{zz}$ and $\tilde{A}_{zx}$ depend explicitly of the droop gains $K_P$ and $K_Q$, stability of \eqref{inv1_red}--\eqref{inv2_red} will largely rely on their chosen values. Given that, we next focus on deriving conditions on the droop gains $K_P$ and $K_Q$ that assure stability of the equilibrium point $(x^*, z^*)=(0_N,0_{2N})$ of \eqref{inv1_red}--\eqref{inv2_red}. 
	\subsection{Fast Boundary-layer Subsystem}
	It is easy to notice that the system \eqref{inv1_red}--\eqref{inv2_red} is in singularly perturbed form with two timescales $t$ and $t_z$. The timescale separation here can be attributed to the time-constant of the inverters' low-pass filters. Once again, we employ singular perturbation \cite{Kokotovic} to analyze this system. We first compute the zeroth-order manifold $z_0(x)=\Gamma_0x$ where $\Gamma_0=(-\tilde{A}_{zz}^{-1}\tilde{A}_{zx})$ and use a change of variables $\eta:=(z-z_0)$ to recover the \textit{fast boundary-layer subsystem} 
	\begin{align}
	\frac{d\eta}{dt_z}=\tilde{A}_{zz} \eta \label{fastboundarydyn},\hspace{7mm}\eta\in\mathbb{R}^{2N}
	\end{align}
	where $\tilde{A}_{zz}:=\tilde{A}_{zz}(K_P,K_Q)\in\mathbb{R}^{2N\times 2N}$.  This system characterizes the dynamics of the frequencies $\omega$ and amplitudes $V$ of the inverters' output voltages when decoupled from the much slower dynamics of the voltage angles $\delta$.
	\subsection{Stability of Fast Boundary-layer Subsystem}
	Through algebraic manipulations, we can obtain the matrix $\tilde{A}_{zz}$ in the \textit{upper triangular} form
	\begin{align}
	\tilde{A}_{zz}:=\begin{pmatrix}-I_{N} & \star\\
	0_{N\times N} & E  \end{pmatrix} \label{Azztilde}
	\end{align}
	with the matrix $E=\{e_{ij}\}_{i,j\in\mathcal{N}}$ being defined  as follows:
	\begin{align}
	e_{ij}=\begin{cases}k_{q,i}\nu_i-1,\hspace{5mm} i=j\\
	k_{q,i}\nu_{ij}, \hspace{10mm}i\neq j\hspace{2mm} \text{and } (j,i)\in\mathcal{E} \\
	0, \hspace{17mm} \text{otherwise}
	\end{cases}	
	\end{align} 
	The terms $\nu_i$ and $\nu_{ij}$ can be expressed as:
	\begin{align}
	\nu_i&= \sum_{k\in\mathcal{E}_i} \Big[\frac{dQ_i}{dI_{D,k}}\Big(\frac{R_k \cos(\delta_i)+X_k\sin(\delta_i)}{R_k^2+X_k^2}\Big)\nonumber\\
	&-\frac{dQ_i}{dI_{Q,k}}\Big(\frac{X_k \cos(\delta_i)-R_k\sin(\delta_i)}{R_k^2+X_k^2}\Big)\Big]\theta_k-\frac{dQ_i}{dV_i}, \hspace{5mm} i\in\mathcal{N}\label{nu_i}
	\end{align}
	\begin{align}
	\nu_{ij}&=\Big[-\frac{dQ_i}{dI_{D,k}}\Big(\frac{R_k \cos(\delta_j)+X_k\sin(\delta_j)}{R_k^2+X_k^2}\Big)\nonumber\\
	&+\frac{dQ_i}{dI_{Q,k}}\Big(\frac{X_k\cos(\delta_j)-R_k\sin(\delta_j)}{R_k^2+X_k^2}\Big)\Big]\theta_{k},\nonumber\\
	&\hspace{45mm} i,j\in\mathcal{N},\; k=(i,j)\in\mathcal{E}
	\end{align}
	where
	\begin{align}
	\theta_k=\begin{cases}1, \text{ if inverter } i \text{ lies at the end of line } k\\
	-1, \text{ if inverter } i \text{ lies in the beginning of line } k\end{cases}\nonumber
	\end{align}
	and $\mathcal{E}_i$ denotes the set of distribution lines connected to inverter $i$.
	The next lemma affirms stability of  \eqref{fastboundarydyn}.
	\begin{lemma}
		The equilibrium $\eta^*=0_{2N}$ of the fast boundary-layer subsystem  \eqref{fastboundarydyn} is asymptotically stable when the voltage droop gains satisfy $K_Q\in\mathcal{A}$ where $\mathcal{A}:=\{K_Q\in\mathbb{R}^N\;\;|\hspace{2mm} E \text{ is Hurwitz } \}$.	\label{lemma2Azztilde}
	\end{lemma}	
It is easy to see that the matrix $E$ is Metzler. A strictly diagonally dominant Metzler matrix is stable; therefore, we can extract from Lemma~\ref{lemma2Azztilde}, the following stability conditions. 
	\begin{lemma}
		Let $\nu_i<0$, for every inverter $i\in\mathcal{N}$, hold. Then, the equilibrium $\eta^*=0_{2N}$ of the subsystem \eqref{fastboundarydyn} is asymptotically stable when the voltage droop gains of the inverters satisfy:
		\begin{align}
		k_{q,i}&>0, &&\text{ if } |\nu_i|>=\sum\limits_{j\in\mathcal{N}_i}|\nu_{ij}|\\
		k_{q,i}&<\frac{1}{\sum\limits_{j\in\mathcal{N}_i}|\nu_{ij}|-|\nu_i|},&&\text{ if } |\nu_i|<\sum\limits_{j\in\mathcal{N}_i}|\nu_{ij}|
		\end{align}
		where $\mathcal{N}_i$ is the set of inverters adjacent to inverter $i$.
		\label{lemmaAzztilde}
	\end{lemma}	
	\par The main implication of the above lemma is that inverters can exploit decentralized stability criteria and tune their voltage droop gains properly using only information from the neighbors in order to assure stability of their fast dynamics. The following corollary is a byproduct of the above lemmas. Its proof can be found in the Appendix.
	\begin{corollary}
		Let $f(x,z):=A_{xz}z,\;\;	g(x,z):=\tilde{A}_{zx}z+\tilde{A}_{zz}z$ and the voltage droop gains $K_Q\in\mathbb{R}^N$ be chosen so that $E$ and thus $\tilde{A}_{zz}$ are Hurwitz. Then, $V_f:=\eta^TH \eta\in\mathbb{R}_{+}$ with $H\succ 0$ is a Lyapunov function for the fast subsystem \eqref{fastboundarydyn} satisfying the following two inequalities
		\begin{align}
		\frac{\partial V_f}{\partial x}g(x,\eta+z_0)&\leq -\alpha_2 \psi_2^2(\eta)\\
		-\frac{\partial V_f}{\partial \eta}f(x,\eta+z_0)&\leq \gamma_1 \psi_2^2(\eta)+\beta_2\psi_1(x)\psi_2(\eta) \label{pertterm}
		\end{align}
		with  $\alpha_2:=\lambda_{min}(\Lambda)$ where $\Lambda:=-(\tilde{A}_{zz}^TH+H\tilde{A}_{zz})$. Further, $\psi_1(x):=\|x\|_2$, $\psi_2(\eta):=\|\eta\|_2$, $\gamma_1:=-\lambda_{min}(Z)$, $Z:=(H\Gamma_0A_{xz}+H^T\Gamma_0A_{xz})$, $\beta_2:=\sigma_{max}(\Theta)$, $\Theta:=Z\Gamma_0$. 
		\label{corollary1}
	\end{corollary}
	\par The key insight here is that stability of the fast dynamics of \eqref{fastboundarydyn} depends only on the voltage droop gains $K_Q$ and is completely independent of the frequency gains $K_P$. 
	\subsection{Slow  Subsystem}
	The slow subsystem can be obtained as
	\begin{align}
	\frac{dx}{dt}=A^{(s)}x, \hspace{5mm} A^{(s)}\in\mathbb{R}^{N\times N} \label{slowreduced}
	\end{align}
	where $A^{(s)}=(-A_{xz}\tilde{A}_{zz}^{-1}\tilde{A}_{zx})$. 
	The following lemma puts forward a stability condition for $A^{(s)}$.
	\begin{lemma}
		The equilibrium $x^*=0_{N}$ of the slow dynamics is asymptotically stable when $K_Q, K_P\in\mathcal{B}$ where $\mathcal{B}:=\{K_Q, K_P\in\mathbb{R}^N\;\;|\hspace{3mm} A^{(s)} \text{ is Hurwitz }\}$.	\label{lemmaAs}
	\end{lemma}	
	The following corollary naturally arises for system \eqref{slowreduced}. Its proof can be found in the Appendix.
	\begin{corollary}
		Let the voltage and frequency droop gains $K_Q$ and $K_P$ be chosen so that $A^{(s)}$ is Hurwitz. Then, $V_s:=x^TPx$ with $P\succ 0$ is a Lyapunov function for the slow  subsystem \eqref{slowreduced} of the reduced inverter-based power grid  satisfying the following two inequalities
		\begin{align}
		\frac{\partial V_s}{\partial x}f(x,z_0)&\leq -\alpha_1 \psi_1^2(x)\\
		\frac{\partial V_s}{\partial x}[f(x,\eta+z_0)-f(x,z_0)]&\leq \beta_1 \psi_1(x)\psi_2(\eta)
		\end{align}
		with $\alpha_1:=\lambda_{min}(Q^{(s)})$, $Q^{(s)}:=-(A^{(s)^T}P+PA^{(s)})$,  $\psi_1(x):=\|x\|_2$, $\psi_2(\eta):=\|\eta\|_2$, $\beta_1:=\sigma_{max}\Big((P+P^T)A_{xz}\Big)$. 
		\label{corollary2}
	\end{corollary}
	Next, we examine stability of the reduced power grid. 
	\subsection{Stability of Reduced-order Inverter-based Power Grid}
We will now explore what other additional condition is required, for stability of the original reduced-order system \eqref{inv1_red}--\eqref{inv2_red}. We start by recasting system \eqref{inv1_red}--\eqref{inv2_red} using the manifold $z_0$ and the change of variables $\eta:=(z-z_0)$  as
	\begin{align}
	\frac{dx}{dt}&=A_{xz}\Gamma_0x+A_{xz}\eta\label{fullinv1}\\
	\varepsilon_1\frac{d\eta}{dt}&=(-\varepsilon_1\Gamma_0A_{xz}\Gamma_0)x+(\tilde{A}_{zz}-\varepsilon_1\Gamma_0A_{xz})\eta  \label{fullinv2}
	\end{align}
 Next, we state Theorem\ref{theorem1} whose proof can be found in the Appendix.
	\begin{theorem}
		Consider the reduced-order inverter-based power grid \eqref{fullinv1}--\eqref{fullinv2}. Let Corollaries~\ref{corollary1} and \ref{corollary2} hold with  $V_f:=\eta^TH \eta$ and $V_s:=x^TPx$ being Lyapunov functions for the fast and slow subsystems, \eqref{fastboundarydyn} and  \eqref{slowreduced}, respectively. Then, $v(x,\eta):=(1-d_1)V_s(x)+d_1V_f(\eta)$ with $d_1=d_1^*=\beta_1/(\beta_1+\beta_2)$ is a Lyapunov function for the system \eqref{fullinv1}--\eqref{fullinv2} and the equilibrium   $(x^*,\eta^*)=(0_N,0_{2N})$ of the system  \eqref{fullinv1}--\eqref{fullinv2} is asymptotically stable when 
		\begin{align}
		\varepsilon_1<\varepsilon_1^*:=\frac{\alpha_1\alpha_2}{\alpha_1\gamma_1+\beta_1\beta_2} \label{eps1bound}
		\end{align}
		where $\alpha_1:=\lambda_{min}(Q^{(s)})$ with $Q^{(s)}:=-(A^{(s)^T}P+PA^{(s)})$ and $\alpha_2:=\lambda_{min}(\Lambda)$ with $\Lambda:=-(\tilde{A}_{zz}^TH+H\tilde{A}_{zz})$. Further, $\beta_1:=\sigma_{max}\Big((P+P^T)A_{xz}\Big)$, $\gamma_1:=-\lambda_{min}(Z)$, $Z:=(H\Gamma_0A_{xz}+H^T\Gamma_0A_{xz})$, $\beta_2:=\sigma_{max}(\Theta)$ and $\Theta:=Z\Gamma_0$. 
		\label{theorem1}
	\end{theorem}
So the reduced-order system \eqref{fullinv1}--\eqref{fullinv2} would be stable when $\varepsilon_1$ respects the upper bound in \eqref{eps1bound} and the voltage droop gains $K_{Q}$ and frequency droop gains $K_{P}$ are chosen so that the fast and slow subsystems \eqref{fastboundarydyn} and \eqref{slowreduced} are both stable. The following corollary (whose proof can be found in the Appendix) stems from Theorem~\ref{theorem1}.
	\begin{corollary}
		Let Theorem 1 hold with $v(x,\eta)$  being a Lyapunov function for the reduced inverter-based power grid \eqref{fullinv1} and \eqref{fullinv2}. Then $v(x,\eta)$ satisfies
		\medmuskip=0mu
		\begin{align}
		&\begin{bmatrix}\frac{\partial V}{\partial x} & \frac{\partial V}{\partial \eta}\end{bmatrix}h(x,\eta,y_0)\leq -\alpha_3 \psi_3^2(x,\eta)\label{thm2bound1}\\
		&\begin{bmatrix}\frac{\partial V}{\partial x} & \frac{\partial V}{\partial \eta}\end{bmatrix}[h(x,\eta,\xi+y_0)-h(x,\eta,\xi+y_0)]\leq \beta_3 \psi_3(x,\eta)\psi_4(\xi)\label{thm2bound3}
		\end{align}
		where $\psi_3(x,\eta):=\|\psi_1(x), \psi_2(\eta)\|_2$, $\psi_4(\xi):=\|\xi\|_2$, $\alpha_3:=\varepsilon_1\lambda_{min}(Q_v)$ and $\beta_3:=\sigma_{max}((\overline{H}+\overline{H}^T)A_{zy})$ with $\overline{H}:=d_1H$. \label{corollary4}
	\end{corollary}
We are now ready to find conditions for stability of the overall system with the electromagnetic network dynamics.
	\subsection{Stability of Full Inverter-based Power Grid}
 We begin by specifying
 \medmuskip=0mu
	\begin{align}
	&h(x,\eta,y):=\nonumber\\
	&\begin{bmatrix}&A_{xz}\varepsilon_1\Gamma_0x+A_{xz}\varepsilon_1\eta\\
	&(A_{xz}+A_{zz}\Gamma_0-\Gamma_0A_{xz}\varepsilon_1\Gamma_0)x+(A_{zz}-\Gamma_0A_{xz}\varepsilon_1)\eta+A_{zy}y
	\end{bmatrix}
	\end{align}
	With the change of variables $\eta=(z-z_0)$ and $\xi:=(y-y_0)$, we can write the full system \eqref{inv1_z}--\eqref{inv3_z} with respect to time-scale $t_z$ as
	\begin{align}
	\frac{dx}{dt_z}&=A_{xz}\varepsilon_1\Gamma_0x+A_{xz}\varepsilon_1\eta\label{full1}\\
	\frac{d\eta}{dt_z}&=(A_{xz}+A_{zz}\Gamma_0-\Gamma_0A_{xz}\varepsilon_1\Gamma_0+A_{zy}(A_{0z}\Gamma_0\nonumber\\&+A_{0x}))x	+(A_{zz}-\Gamma_0A_{xz}\varepsilon_1+A_{zy}A_{0z})\eta+A_{zy}\xi\label{full2}\\
	\varepsilon_3\frac{d\xi}{dt_z}&=\tilde{A}_{yy} \xi-\varepsilon_3A_{0z}\frac{d\eta}{dt_z}-\varepsilon_3(A_{0x}+A_{0z}\Gamma_0)\frac{dx}{dt_z}\label{full3}
	\end{align}
	Clearly, our final goal is to find conditions for stability of \eqref{full1}--\eqref{full3}. We designate
	\begin{align}
	\mu(x,\eta,y):=D_2[(A_{yx}+A_{yz}\Gamma_0)x+A_{yz}\eta+A_{yy}y]\in\mathbb{R}^{2M}\nonumber
	\end{align}
	and compute $\mu(x,\eta,\xi+y_0):=\tilde{A}_{yy}\xi$. By combining the stability properties of the decoupled reduced-order power grid and the network dynamics, we arrive at our final result.
	\begin{theorem}
		Consider the full inverter-based power grid \eqref{full1}--\eqref{full3}. Let Corollaries~\ref{corollary3} and \ref{corollary4} hold with  $W:=\xi^TP_{\xi} \xi$ being a Lyapunov function for the decoupled electromagnetic network dynamics \eqref{veryfastsystem} and $v(x,\eta)$  a Lyapunov function for the reduced-order power grid \eqref{fullinv1}--\eqref{fullinv2}. Then, $\mathcal{V}(x,\eta,\xi):=(1-d_2)v(x,\eta)+d_2W(\xi)$ with $d_2:=\beta_3/(\beta_3+\beta_4)$ is a Lyapunov function for the system \eqref{full1}--\eqref{full3} and the equilibrium   
		$(x^*,\eta^*,\xi^*)=(0_N,0_{2N},0_{2M})$ of the full inverter-based power grid \eqref{full1}--\eqref{full3} is asymptotically stable as long as
		\begin{align}
		\varepsilon_3:=\frac{\Big(\frac{L_1}{\omega_b}\cdot\cdot\cdot\frac{L_M}{\omega_b}\Big)^{1/M}}{T_P}<	\varepsilon_3^*:=\frac{\alpha_3\alpha_4}{\alpha_3\gamma_2+\beta_3\beta_4}	\label{e3_expression}	\end{align}
	 Further, $\alpha_3:=\varepsilon_1\lambda_{min}(Q_v)$, $\alpha_4:=\lambda_{min}(Q_{\xi})$ where $Q_{\xi}:=-(\tilde{A}_{yy}^TP_{\xi}+P_{\xi}\tilde{A}_{yy})$, $\beta_3:=\sigma_{max}((\overline{H}+\overline{H}^T)A_{zy})$, $\overline{H}:=d_1H$, $\beta_4:=\max(\sigma_{max}(\Theta_{\xi x}),\sigma_{max}(\Theta_{\xi \eta}))$, $\gamma_2:=-\lambda_{min}(\Theta_{\xi\xi})$ with $\Theta_{\xi\xi}$ given by \eqref{thetaxixi}.
		\label{theorem2}
	\end{theorem}	
	\par Essentially, this theorem says that we have stability of the original inverter-based power grid \eqref{full1}--\eqref{full3} when: a) the reduced-order approximated system obtained by suppressing the network dynamics is stable (Theorem~\ref{theorem1}) and, b) the parameter $\varepsilon_3$ respects the upper bound $\varepsilon_{3}^*$. 
	\par Our analytical framework can be a useful tool for systematic and efficient stability analysis of  inverter-based power grids. The traditional approach to stability analysis usually involves adjusting the droop gains $K_P$ and $K_Q$ and repeatedly computing the eigenvalues of the $(3N+2M)$-dimensional full system matrix  until this becomes Hurwitz.   With our framework, one has to only assure that the $N$-dimensional matrices $E$ and $A^{(s)}$ are Hurwitz and the parameters $\varepsilon_1$ and $\varepsilon_3$ respect some well-defined bounds. 
	
	\section{Numerical Validation}
	We corroborate our theoretical results numerically and illustrate how our framework can be practically implemented via an example on the IEEE 13-bus distribution grid.
	\subsection{Set-up}
	We use the standard IEEE 13-bus test feeder \cite{13bussystem} model to validate our results which we modify by placing a single inverter at each bus. Further, we reduce four buses of the original system to two buses in the modified system. Each inverter $i$, where $i\in\mathcal{N}$, has a power rating $S_{n,i}=10$kVA and low-pass filter time-constant $T_{P,i}=T_{Q,i}=0.0318$s \cite{turitsyn2}.  Here, $\mathcal{N}$ denotes the set of the 10 inverters.  In our example we choose $V_b=381.58$V as the base voltage, $S_b=10$kVA as the base power and compute the base impedance as $Z_b=(V_b^2/S_b)=14.56\Omega$. 
	\subsection{Implementation of the Proposed Framework}
\begin{center}
		\begin{figure}[h!]
			\includegraphics[scale=0.34]{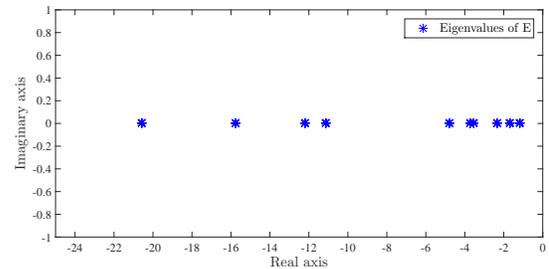}
			\caption{Eigenvalues of the matrix E with $k_{q,i}=0.05,\;\forall i\in\mathcal{N}$.}
			\label{eigE}
		\end{figure}	
	\end{center}

 By applying the conditions in Lemma~\ref{lemmaAzztilde} we arrive at the following inequalities $k_{q,8}<0.2,\;\; k_{q,i}>0,\hspace{3mm}\forall i\in\mathcal{N}\setminus \{8\}$. Choosing $k_{q,i}=0.05$ yields the eigenvalues of $E$ shown in Fig.~\ref{eigE}. Although these inequalities are not very restrictive for the voltage droop gains $K_Q$, recall that, these droop gains have to still be chosen carefully, in conjuction with the frequency droop gains $K_P$, so that $A^{(s)}$ is Hurwitz.
\vspace{-7mm}
	\begin{center}
		\begin{figure}[h!]
			\includegraphics[scale=0.34]{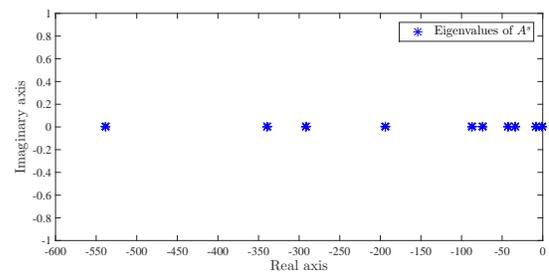}
			\caption{Eigenvalues of the matrix $A^{(s)}$ with $k_{q,i}=0.05,k_{p,i}=0.6,\;\forall i\in\mathcal{N}$. The largest eigenvalue is $-0.893$.}
			\label{eigAs}
		\end{figure}
	\end{center}
\vspace{-10mm}
	\par Having chosen the voltage droop gains $k_{q,i}$, we then compute $A^{(s)}$ symbolically in terms of the frequency droop gains $K_P$. We now have to choose frequency droop gains that yield a Hurwitz matrix $A^{(s)}$. One suitable choice for these gains is $k_{p,i}=0.6, \forall i\in\mathcal{N}$, as can be coroborrated by Fig.~\ref{eigAs}. The largest eigenvalue of $A^{(s)}$ is $-0.893$. To proceed, we construct Lyapunov functions $V_s=x^TPx$ and $V_f=\eta^T H\eta$ for the slow and fast subsystems by trivially choosing $Q^{(s)}=I_{N}$ and $\Lambda=I_{2N}$ where $A^{(s)^T}P+PA^{(s)}=-Q^{(s)}$ and $\tilde{A}_{zz}^TH+H\tilde{A}_{zz}=-\Lambda$.  With these Lyapunov functions, we obtain through Theorem~\ref{theorem1} the following upper bound for the time-constant of the inverters' low-pass filters $\varepsilon_1^*=0.0178\cdot 10^{-5}$. 	This bound reflects the maximum time constant of inverters' low-pass filters for which stability of $E$ and $A^{(s)}$ readily translates into stability of the reduced-order inverter-based power grid. For the particular Lyapunov functions that we chose here, this bound turns out to be quite convervative as the value of $\varepsilon_1$, which corresponds to the time-constants $T_{P,i}, T_{Q,i}$, is $0.0318$. One has to experiment with different Lyapunov functions in order to arrive at a less conservative bound. This is beyond the scope of this work, but could be considered in future work.
	\par We now focus on the decoupled network dynamics which are always stable. We let $Q_{\xi}=I_{2M}$ where	$\tilde{A}_{yy}^T P_{\xi}+P_{\xi}\tilde{A}_{yy}=-Q_{\xi}$ and construct a Lyapunov function $W:=\xi^TP_\xi \xi$ for these dynamics. We let $\varepsilon_1=\varepsilon_1^{*}/2$ and use Theorem~\ref{theorem2} to obtain $\varepsilon_3^*=7.84\cdot 10^{-10}$. As the actual value of $\varepsilon_3$, which can be computed using \eqref{e3_expression}, is $6.8705\cdot 10^{-4}$ we see that this bound is also quite conservative. 
		\par
\begin{center}
		\begin{figure}
			\includegraphics[scale=0.34]{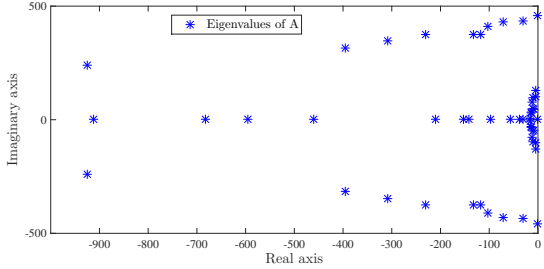}
			\caption{Eigenvalues of full matrix A with $k_{q,i}=0.05,k_{p,i}=0.6,\;\forall i$. The eigenvalues which the largest real part are  $-0.918+456.814i, -0.918-456.814i,-0.922$.}
			\label{eigA}
		\end{figure}
	\end{center}

 Overall, our numerical example shows that the bounds $\varepsilon_1^*$ and $\varepsilon_3^*$ can be quite conservative for certain  Lyapunov functions. One could try to come up with ``better" Lyapunov functions that would lead to less conservative upper bounds.  As the conditions are only sufficient, the full system may still be stable even when they are not met. To illustrate this, we compute  the eigenvalues of the full system matrix $A$ with the chosen droop gains, $k_{q,i}=0.05$ and $k_{p,i}=0.6$. As can be seen from Fig.~\ref{eigA},  all eigenvalues of $A$ are negative despite the fact that the bounds $\varepsilon_1^*$ and $\varepsilon_3^*$ are not respected. 
	\vspace{-10mm}	
	\section{Conclusions}
	We studied the stability problem for distribution grids with droop-controlled inverters and electromagnetic network dynamics. We presented a systematic framework that builds on singular perturbation and Lyapunov theories for carrying out  stability analysis in a compositional and efficient manner. By deploying our framework, we derived sufficient stability conditions for the full system. Our theoretical results are numerically corroborated via an example on the IEEE 13-bus distribution grid.

	\nocite{*}
	\IEEEpeerreviewmaketitle
	\bibliographystyle{unsrt}
	\bibliography{CDC20_Stability_of_DGs}{}
	\thanks
	
	\appendix
	\textit{Proof of Corollary~\ref{corollary1}}
		\begin{proof}
		When $\tilde{A}_{zz}$ is Hurwitz, along the trajectories of the fast boundary-layer subsystem \eqref{fastboundarydyn} we have
		\begin{align}
		&\frac{\partial V_f}{\partial \eta}g(x,\eta+z_0)=\eta^T(\tilde{A}_{zz}^T H+H\tilde{A}_{zz})\eta\nonumber\\
		&=-\eta^T\Lambda \eta\leq -\lambda_{min}(\Lambda)\|\eta\|_2^2=-\alpha_2\psi_2^2(\eta) \label{bound2}
		\end{align}
		where $\alpha_2:=\lambda_{min}(\Lambda)$, $\psi_2(\eta):=\|\eta\|_2$ and $\Lambda\succ 0$. With this, we conclude that $V_f$ is a Lyapunov function for the fast system \eqref{fastboundarydyn}. We are left to show that the perturbation term in \eqref{pertterm} is bounded appropriately. Expanding this term yields
		\begin{align}
		\Big(&-\frac{\partial V_f}{\partial y}\frac{\partial z_0}{\partial x}\Big)f(x,\eta+z_0)=-\eta^T Z \eta -\eta^T\Theta x \nonumber\\
		&\leq -\lambda_{min}(Z)\|\eta\|_2^2+\sigma_{max}(\Theta)\|\eta\|_2\|x\|_2\nonumber\\
		&=\gamma_1 \psi_2^2(\eta)+\beta_2\psi_1(x)\psi_2(\eta) \label{bound4}
		\end{align}
		where  $\psi_1(x):=\|x\|_2$, $\gamma_1:=-\lambda_{min}(Z)$ and $Z:=(H\Gamma_0A_{xz}+H^T\Gamma_0A_{xz})$. Further, $\beta_2:=\sigma_{max}(\Theta)$ where $\Theta:=Z\Gamma_0$ and with that we  complete the proof.
	\end{proof}
		\textit{Proof of Corollary~\ref{corollary2}}
		\begin{proof}
		With $A^{(s)}$ being Hurwitz, along the trajectories of the slow subsystem \eqref{slowreduced} we obtain
		\begin{align}
		\frac{\partial V_s}{\partial x}f(x,z_0)&=x^T(A^{(s)^T}P+PA^{(s)})\leq-\alpha_1\psi_1^2(x) \label{bound1}
		\end{align}
		where $\alpha_1:=\lambda_{min}(Q^{(s)})$, $\psi_1(x):=\|x\|_2$ and $Q^{(s)}\succ 0$. With this, we deduce that $V_s$ is a Lyapunov function for the slow system \eqref{slowreduced}. Eventually, we can also bound the term:
		\begin{align}
		&\frac{\partial V_s}{\partial x}\Big[f(x,\eta+z_0(x))-f(x,z_0(x))\Big]=x^T(P+P^T)A_{xz}\eta\nonumber\\
		&\leq \sigma_{max}\Big((P+P^T)A_{xz}\Big)\|x\|_2\|\eta\|_2=\beta_1 \psi_1(x) \psi_2(\eta) \label{bound3}
		\end{align}
		where $\beta_1:=\sigma_{max}\Big((P+P^T)A_{xz}\Big)$, $\psi_2(\eta):=\|\eta\|_2$ and conclude the proof.
	\end{proof}	
\vspace{-1mm}
		\textit{Proof of Theorem~\ref{theorem1}}
	\begin{proof}
		We construct a composite candidate Lyapunov function for the system \eqref{fullinv1}-\eqref{fullinv2} $v(x,\eta):=(1-d_1)V_s(x)+d_1V_f(\eta)$.	Calculating the derivative of $v(x,\eta)$ along the trajectories of the full system \eqref{fullinv1}-\eqref{fullinv2}, we obtain
		\begin{align}
		\frac{dv}{dt}&=(1-d_1)\frac{\partial V_s}{\partial x}f(x,z_0)+(1-d_1)\frac{\partial V_s}{\partial x}\Big[f(x,\eta+z_0)\nonumber\\
		&-f(x,z_0)\Big]+\frac{d_1}{\varepsilon_1}\frac{\partial V_f}{\partial \eta}g(x, \eta+z_0)-d_1\frac{\partial V_f}{\partial \eta} \frac{\partial z_0}{\partial x} f(x,\eta+z_0) \nonumber
		\end{align}
		Applying the inequalities  \eqref{bound2}, \eqref{bound4}, \eqref{bound1}, \eqref{bound3} yields
		\begin{align}
		\frac{dv}{dt}&\leq -(1-d_1)\alpha_1 \psi_1^2(x)-\frac{d_1}{\varepsilon_1}\alpha_2 \psi_2^2(\eta)\nonumber\\
		&+(1-d_1)\beta_1 \psi_1(x)\psi_2(\eta)+d_1\gamma_1\psi_2^2(\eta)+d_1\beta_2 \psi_1(x)\psi_2(\eta)
		\end{align}
		which can be expressed in quadratic form as
		\thinmuskip=0mu
		\begin{align}
		\setlength\arraycolsep{0pt}
		&\frac{dv}{dt}\leq -\begin{bmatrix} \psi_1(x) \\ \psi_2(\eta)\end{bmatrix}^TQ_v\begin{bmatrix}\psi_1(x)\\ \psi_2(\eta)\end{bmatrix} \label{vdot}
		\end{align}
		where
		\begin{align}
		Q_v:=\begin{bmatrix}
		(1-d_1)\alpha_1 & -\frac{1}{2}(1-d_1)\beta_1-\frac{1}{2}d_1\beta_2\\
		-\frac{1}{2}(1-d_1)\beta_1-\frac{1}{2}d_1\beta_2 & d_1(\alpha_2/\varepsilon_1-\gamma_1)
		\end{bmatrix}
		\end{align}
		Positive definiteness of $Q_v$ is guaranteed when
		\begin{align}
		\varepsilon_1 \leq \underbrace{\frac{\alpha_1\alpha_2}{\alpha_1\gamma_1+\frac{1}{4d_1(1-d_1)}[(1-d_1)\beta_1+d_1\beta_2]^2}}_{:=\varepsilon_{1,d}}.
		\end{align}
		Choosing $d_1^*:=\beta_1/(\beta_1+\beta_2)$, yields the maximum value of $\varepsilon_1^{*}=\frac{\alpha_1\alpha_2}{\alpha_1\gamma_1+\beta_1\beta_2}$. That, completes the proof.
	\end{proof}
	\textit{Proof of Corollary~\ref{corollary3}}
		\begin{proof}
		We specify $\rho:=[\psi_1(x),\;\; \psi_2(\eta)]^T$, $\psi_3(x,\eta):=\|\rho\|_2$ and $\psi_4(\xi):=\|\xi\|_2$. Let $W(\xi):=\xi^TP_{\xi}\xi$ be a candidate Lyapunov function for the very fast dynamics \eqref{veryfastsystem}. The derivative of $W$ along the trajectories of \eqref{veryfastsystem} is
		\begin{align}
		\frac{\partial W}{\partial \xi}\mu(x,\eta,\xi+y_0)&:=\xi^T(\tilde{A}_{yy}^TP_{\xi}+P_{\xi}\tilde{A}_{yy})\xi\nonumber\\
		&\leq -\lambda_{min}(Q_{\xi})\|\xi\|_2^2= -\alpha_4 \psi_4^2(\xi)\label{fullbound2}
		\end{align}
		where, as $Q_{\xi}\succ 0$ (due to $\tilde{A}_{yy}$ being Hurwitz) we conclude that $W$ is a Lyapunov function for  \eqref{veryfastsystem}. Finally, we also have
		\begin{align}
		&-\frac{\partial W}{\partial \xi}\begin{bmatrix}\frac{\partial y_0}{\partial x} & \frac{\partial y_0}{\partial \eta} \end{bmatrix}h(x,\eta,\xi+y_0)\nonumber\\
		&=-\xi^T \Theta_{\xi x}x-\xi^T \Theta_{\xi\eta}\eta-\xi^T \Theta_{\xi \xi}\xi\nonumber\\
		&\leq \max (\sigma_{max}(\Theta_{\xi x}),\sigma_{max}(\Theta_{\xi \eta}))\psi_4(\xi) \sqrt{2} \psi_3(x,\eta)\nonumber\\
		& -\lambda_{min}(\Theta_{\xi \xi})\psi_4^2(\xi):= \beta_4\psi_3(x,\eta)\psi_4(\xi)+\gamma_2\psi_4^2(\xi)\label{fullbound4}
		\end{align}
		where $\beta_4:=\max (\sigma_{max}(\Theta_{\xi x}),\sigma_{max}(\Theta_{\xi \eta}))\sqrt{2}$ and $\gamma_2:=-\lambda_{min}(\Theta_{\xi \xi})$. We note that the matrices 
		$\Theta_{\xi x}, \Theta_{\xi \eta}, \Theta_{\xi \xi}$ are as specified in \eqref{thetaxichi}, \eqref{thetaxieta}, \eqref{thetaxixi} and complete the proof.	
	\end{proof}	
	\textit{Proof of Corollary~\ref{corollary4}}
		\begin{proof}
		From \eqref{vdot}, we readily have that
		\begin{align}
		\begin{bmatrix}\frac{\partial v}{\partial x} & \frac{\partial v}{\partial n}\end{bmatrix}h(x,\eta,y_0)\leq-\varepsilon_1\lambda_{min}(Q_v)\psi_3^2(x,\eta)=-\alpha_3\psi_3^2(x,\eta) \label{fullbound1}
		\end{align}
		where $\alpha_3:=\varepsilon_1\lambda_{min}(Q_v)$.		By letting $\overline{P}:=(1-d_1)P$ and $\overline{H}:=d_1H$ one can express $v(x,\eta)$ as $v(x,\eta):=x^T\overline{P}x+\eta^T \overline{H}\eta$ and derive the following inequality
		\begin{align}
		&\begin{bmatrix}\frac{\partial V}{\partial x} & \frac{\partial V}{\partial \eta}\end{bmatrix}[h(x,\eta,\xi+y_0)-h(x,\eta,y_0)]\nonumber\\
			&=\eta^T (\overline{H}+\overline{H}^T)A_{zy}\xi\leq \sigma_{max}\Big((\overline{H}+\overline{H}^T)A_{zy}\Big)\|\eta\|_2\|\xi\|_2\nonumber\\
		&\leq\beta_3\psi_3(x,\eta)\psi_4(\xi)\label{fullbound3}
		\end{align}	
		With that, we complete the proof.
	\end{proof}	
		\vspace{1mm}
	\textit{Proof of Theorem~\ref{theorem2}}
		\begin{proof}
	Similarly as before, we can employ a candidate Lyapunov  $\mathcal{V}(x,\eta,\xi):=(1-d_2)v(x,\eta)+d_2W(\xi)$ and use the inequalities \eqref{fullbound2}, \eqref{fullbound4}, \eqref{fullbound1}, \eqref{fullbound3} to finally obtain $\varepsilon_3^{*}=\frac{\alpha_3\alpha_4}{\alpha_3\gamma_2+\beta_3\beta_4}$ where  $d_2:=\beta_3/(\beta_3+\beta_4)$. That, completes the proof.
	\end{proof}

\end{document}